\documentclass[11pt,preprint]{aastex}

\usepackage{graphicx}
\def\sgr{Sgr~A$^*$}
\def\msun{M_\odot} 
\def\Em{{\mathcal M}} 
\def\nughz{\nu_{\rm GHz}}

\begin{document}

\title{Radio Synchrotron Emission from a Bow Shock around the Gas 
Cloud G2 Heading toward the Galactic Center}

\author{Ramesh Narayan$^1$, Feryal \"Ozel$^2$ and Lorenzo Sironi$^{1,3}$}

\affil{$^1$ Harvard-Smithsonian Center for Astrophysics, 
60 Garden St., Cambridge, MA 02138} 
\affil{$^2$Department of Astronomy, University of Arizona, 933 N. 
Cherry Ave., Tucson, AZ 85721} 
\affil{$^3$NASA Einstein Post-Doctoral Fellow} 

\begin{abstract}

A dense ionized cloud of gas has been recently discovered to be moving
directly toward the supermassive black hole, \sgr, at the Galactic
Center.  In June 2013, at the pericenter of its highly eccentric
orbit, the cloud will be approximately 3100 Schwarzschild radii from
the black hole and will move supersonically through the ambient hot
gas with a velocity of $v_p \approx 5400$~km~s$^{-1}$. A bow shock is
likely to form in front of the cloud and could accelerate electrons to
relativistic energies. We estimate via particle-in-cell simulations
the energy distribution of the accelerated electrons and show that the
non-thermal synchrotron emission from these electrons might exceed the
quiescent radio emission from \sgr\ by a factor of several.  The
enhanced radio emission should be detectable at GHz and higher
frequencies around the time of pericentric passage and in the
following months. The bow shock emission is expected to be displaced
from the quiescent radio emission of \sgr\ by $\sim 33$~mas.
Interferometric observations could resolve potential changes in the
radio image of \sgr\ at wavelengths $\lesssim 6$~cm.

\end{abstract}


\section{Introduction}

Recent sub-millimeter observations revealed a dense ionized cloud of
gas known as G2, rapidly approaching \sgr, the black hole at the
Galactic Center (Gillessen et al.\ 2012). The cloud is on a highly
eccentric trajectory, with a 2011 distance from the black hole of
$1.8\times10^{16}$\,cm.  The pericentric passage, which is expected
to occur in mid 2013, will bring the cloud within $4\times10^{15}$\,cm
from the supermassive black hole. Given the mass of the black hole,
which is determined through observations of nearby stellar orbits to
be $M=4.3\times10^6\;\msun$ (Ghez et al.\ 2008; Gillessen et
al.\ 2009), this pericentric distance is only $R_p=3100 \,R_S$, where
the Schwarzschild radius $R_S=1.27\times10^{12}$\,cm.

The accretion flow around the black hole extends to the Bondi radius
$\sim10^5R_S$ (e.g., Yuan, Quataert \& Narayan 2003) and powers the
multiwavelength emission observed from it. The flux at 1~GHz is
$\simeq 0.5$~Jy, rising to $\approx 4$~Jy at 500~GHz, before rapidly
declining at higher frequencies. The radio emission has been
successfully modeled as synchrotron radiation from relativistic
electrons, either in a radiatively inefficient accretion flow (ADAF)
(Narayan, Yi, \& Mahadevan 1995; \"Ozel, Psaltis \& Narayan 2000) or
in a jet (Falcke \& Markoff 2000). At the lowest end of the spectrum
--- $\nu \sim 1-10$\,GHz --- the radio emission shows flux variability
on a time scale of months to years with a root mean square amplitude
$\sim10\%$, (Zhao et al. 1989; Falcke 1999; Macquart \& Bower 2006).

At its pericentric passage, the gas cloud will interact with the
accretion flow, and this may significantly change its dynamics. Here,
we show that a bow shock is likely to develop as the cloud plows
through the hot, tenuous plasma at $R_p$. In \S2 we calculate the
properties of the bow shock and in \S3 we estimate the energy
distribution of electrons accelerated at this shock. In \S4 we
calculate the extra radio emission that will result from these
accelerated electrons. For likely electron energy distributions, the
additional emission is $\sim10$\,Jy at frequencies $\sim1-10$~GHz.
This is well above the quiescent emission from \sgr\ and should be
easily detectable.  We also show that the flux increase will be
accompanied by significant changes in the spectral index. In \S5 we
summarize our findings and argue that interferometric observations
could resolve potential changes in the radio image of \sgr\ caused by
the interaction of the cloud with the accretion flow.

\section{The Bow Shock around G2 at Pericenter}

We first calculate the dynamics of the interaction of the cloud G2
with the accretion flow around the black hole.  Since the orbit of G2
is highly eccentric, its velocity at pericenter will be
\begin{equation}
v_p \approx \left(\frac{2GM}{R_p}\right)^{1/2}=5400~{\rm km\,s^{-1}}.
\end{equation}
The properties of the ambient gas at the pericentric distance $R_p$
can be obtained from ADAF models of Sgr A$^*$. Within that context and
using Yuan et al. (2003) and Xu et al.\ (2006) as a
guide, Gillessen et al.\ (2012) estimated the gas density and
temperature at $R=R_p$ to be
\footnote{Some models suggest a temperature closer to $10^{8.5}$\,K,
  but this small uncertainty is not important for what follows.}
\begin{equation}
n \approx 930\left(\frac{1.4\times10^4\;R_S}{R_p}\right) ~{\rm cm^{-3}}
=4200 ~{\rm cm^{-3}}, \qquad T \approx 10^{9}\,{\rm K}.
\end{equation}
To estimate the sound speed, we use the fact that the gas in the ADAF
is likely to have its Bernoulli parameter close to zero (Narayan \& Yi
1994). Thus,
\begin{equation}
Be = -\frac{GM}{R_p} +\frac{1}{2}v_R^2+\frac{1}{2}v_\phi^2+w \approx
0,
\end{equation}
where $w=\Gamma p/(\Gamma-1)\rho$ is the enthalpy per unit mass,
$\Gamma=5/3$ is the adiabatic index, $p$ is the pressure, and $\rho$
the density of the gas. Because the radial velocity $v_R$ is small
compared to the azimuthal velocity $v_\phi$, we ignore it.  We also
assume that $v_\phi^2$ is roughly half the Keplerian value, which is
appropriate for an ADAF. Under these assumptions, we estimate the
adiabatic sound speed of the gas, $c_{\rm ad} = \sqrt{\Gamma p/\rho}$,
to be
\begin{equation}
c_{\rm ad} \approx v_p/2 \approx 2700~{\rm km\,s^{-1}}.
\end{equation}
The gas pressure is then given by
\begin{equation}
p=\frac{1}{\Gamma}\rho c_{\rm ad}^2 \approx 3.1\times 10^{-4}~{\rm
erg\,cm^{-3}}.
\end{equation}
Finally, if we assume that the magnetic pressure is about 10\% of the
gas pressure, as is typical in an ADAF, then we estimate the magnetic
field strength in the ambient gas to be $\approx 0.03$\,G.

There is currently no information on the relative orientation of the
orbits of G2 and the ambient gas in the accretion flow. However,
because the velocity $v_p$ of the cloud at pericenter is substantially
larger than the local gas velocity $v_\phi$, we expect the relative
velocity at $R_p$ to be dominated by $v_p$. Thus, G2 will move with a
Mach number $\Em \approx 2$ near pericenter. Using Rankine-Hugoniot
jump conditions for a non-relativistic shock, we estimate for the
shocked gas,
\begin{eqnarray}
\frac{n_{\rm shock}}{n} &=& \frac{(\Gamma+1)\Em^2}{(\Gamma-1)\Em^2+2} 
\approx 2.3, \\
\frac{p_{\rm shock}}{p} &=& \frac{(\Gamma+1)+2\Gamma(\Em^2-1)}{(\Gamma+1)}
\approx 4.8.
\end{eqnarray}

For a Mach number $\Em \approx 2$, the magnetic field strength will be
roughly doubled by compression in the shock, i.e., $B_{\rm
shock}\approx0.06$\,G. Similarly, the post-shock temperature will be a
factor of a few larger than the temperature of the pre-shock gas.
Thus, the mean thermal energy of the shocked electrons should be close
to $m_ec^2$, i.e., the electrons will be quasi-relativistic.

Gillessen et al.\ (2012) estimated the equivalent spherical size of
the cloud in 2011 to be 15 mas, which corresponds to a physical radius
of $1.9 \times 10^{15}$~cm. As the cloud approaches the pericenter
along its highly eccentric orbit, it is expected to be tidally
stretched (Gillessen et al. 2012; Burkert et al.\ 2012), primarily
along the direction of motion. We estimate the cross-sectional area
$A$ of the bow shock using the frontal size of the cloud, which will
remain approximately constant at $\pi (10^{15}{\rm cm})^2$.

\begin{figure}[t]
\centering
   \includegraphics[scale=0.535]{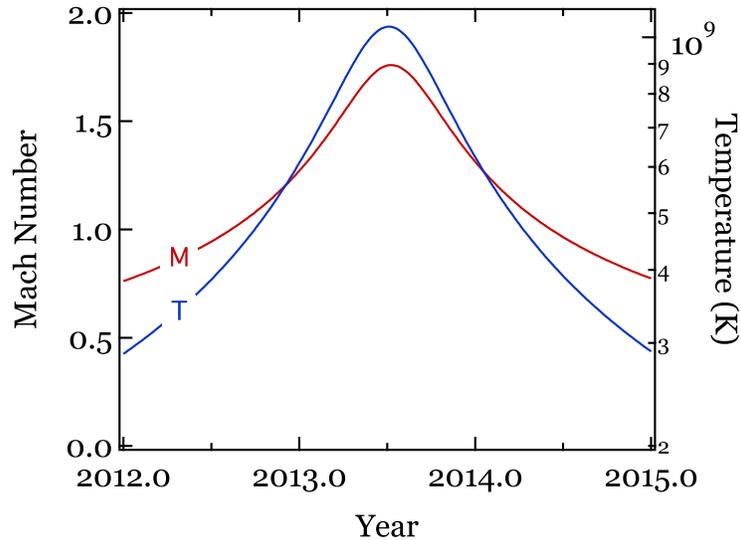}
\caption{Mach number $\Em$ of G2's motion through the hot accretion
  flow around \sgr\ and temperature $T$ of this medium as a function
  of time. Pericentric passage occurs in year 2013.5. Both $\Em$ and
  $T$ peak sharply around pericenter, and a bow shock should be
  present for approximately 6 months on either side of
  pericenter. Since the orientation of the angular momentum vector of
  the accreting gas is not known, for simplicity the gas has been
  assumed to be at rest.}  \mbox{}
\label{fig:mach} 
\end{figure}

Figure \ref{fig:mach} shows the Mach number of G2's motion through the
ambient medium (assuming that the medium is at rest) and the
temperature of the medium as a function of time. A bow shock will form
when $\Em$ exceeds unity, which is expected to last for about a year
around pericenter. The temperature of the medium, which has a
significant effect on particle acceleration (see \S3), is also strongly
peaked around pericenter. Taking the relative velocity between G2 and
the ambient medium to be equal to $v_p$, and conservatively taking the
duration of the bow shock to be 6 months, we estimate the total number
of shocked electrons to be
\begin{equation}
N_{\rm shock} \approx A\;v_{ p}\;t_{ p}\;n = 1.1\times10^{50}. 
\label{eq:Nsh}
\end{equation}

\section{Acceleration of Electrons in the Bow Shock}

A variety of astrophysical evidence suggests that particles are
accelerated efficiently via the Fermi and shock-drift acceleration
mechanisms in shocks (e.g., Blandford \& Eichler 1987). Typically,
these processes give rise to a non-thermal power-law tail in the
energy distribution of the particles.  The parameters of the bow shock
of G2 correspond to an interesting regime that has not been well
studied. The shock is non-relativistic, but the upstream electrons are
quasi-relativistic ($kT_e \lesssim m_ec^2$). In addition, the shock
has a modest Mach number $\Em\approx2$, and the upstream gas is fairly
strongly magnetized, corresponding to an Alfvenic Mach number
$\Em_A\approx8$.

We studied the acceleration of electrons in the bow shock of G2 by
means of two-dimensional first-principles numerical simulations, with
the particle-in-cell (PIC) code TRISTAN-MP (Spitkovsky 2005). The
simulation setup parallels very closely the one employed by Riquelme
\& Spitkovsky (2011), with the magnetic field lying initially in the
simulation plane, oriented at an oblique angle with respect to the
flow velocity. For computational convenience, we chose a reduced mass
ratio $m_p/m_e=100$, but we tested that our results remain the same
for larger mass ratios, when all the physical quantities are scaled
appropriately. Specifically, we ran simulations spanning the range
$m_p/m_e=25-400$, fixing the electron temperature (equal to the proton
temperature) and the shock sonic and Alfvenic Mach numbers, and we
measured the time in units of the inverse proton cyclotron frequency
$\omega_{ci}^{-1}$. We also checked the convergence of our results
with respect to the spatial resolution and the number of computational
particles per cell.

We find that at the shock, a fraction of the incoming electrons are
reflected backward by the shock-compressed magnetic field (Matsukiyo
et al. 2011) or by scattering off of electron whistler waves excited
in the shock transition layer (Riquelme \& Spitkovsky 2011). For
quasi-relativistic electron temperatures, the reflected electrons are
fast enough to remain ahead of the shock, resisting advection
downstream by the oblique pre-shock field. While the electrons gyrate
around the shock, they are energized by shock-drift acceleration
(e.g., Begelman \& Kirk 1990) and form a local non-thermal population,
just upstream of the shock.

\begin{figure}[t]
\centering
   \includegraphics[scale=0.535]{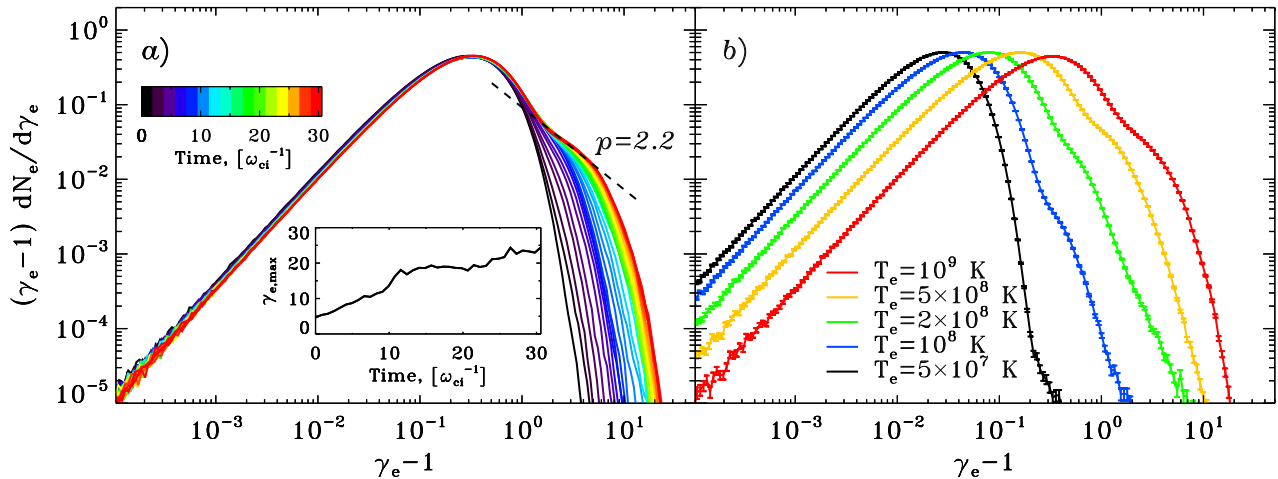}
\caption{Electron energy spectrum just upstream of the shock,
  normalized to the pre-shock electron density. (Left) The temporal
  evolution of the spectrum, for $T_e=10^9$~K, $\Em=2$ and
  $\Em_A=8$. With time, the non-thermal component approaches a
  power-law tail with index $p=2.2$ (dashed line) and its cutoff
  energy steadily increases (insert). (Right) Electron energy spectrum
  at $\omega_{ci}t=14$ for different upstream temperatures (therefore,
  different $\Em$), with fixed $\Em_A=8$ and fixed flow velocity. The
  regime $10^{8.5}\;{\rm K}\lesssim T_e\lesssim 10^9\;{\rm K}$ is relevant  for the accretion flow around \sgr\ at a distance $R_p$ from the black
  hole.}  \mbox{}
\label{fig:time} 
\end{figure}

The temporal evolution of this population is shown in the left panel
of Figure~\ref{fig:time}, for parameters relevant to the bow shock of
G2: $T_p=T_e=10^9 {\rm K}\simeq0.2\, m_ec^2/k$, $\Em=2$ and $\Em_A=8$. At
late times, the shock-accelerated electrons populate a power-law tail
with a slope of $p=2.2$ beyond an electron Lorentz factor
$\gamma_e\simeq2$, containing roughly $5\%$ of the incoming
electrons. The upper energy cutoff of the electron spectrum steadily
increases with time, as shown in the left panel insert of
Figure~\ref{fig:time}, suggesting that the distribution will asymptote
at late times to a power-law with $p\lesssim 2.2$ extending to very
large values of $\gamma_e$. The counter streaming between the incoming
flow and the shock-reflected electrons triggers the Weibel
filamentation instability ahead of the shock (Weibel 1959, Medvedev \&
Loeb 1999), with the wavevector perpendicular to the pre-shock
field.\footnote{We point out that the Weibel mode can only be captured
by means of multi-dimensional simulations and was, therefore, absent
in the one-dimensional experiments of Matsukiyo et al. (2011).} By
scattering off of the magnetic field generated by the Weibel
instability, the shock-reflected electrons are deflected back toward
the shock, participating in a Fermi-like acceleration process. In the
downstream region, they populate a power-law tail of similar
normalization and slope as the pre-shock spectrum shown in the left
panel of Figure~\ref{fig:time}.

The distribution of non-thermal electrons is sensitive to the electron
temperature ahead of the shock. As shown in the right panel of
Figure~\ref{fig:time}, the normalization of the power-law tail is
reduced by almost one order of magnitude when the electron temperature
decreases from $T_e=10^9 K$ (red curve) down to $T_e=10^8 K$ (blue
curve), with the flow velocity staying fixed. If the upstream plasma
is colder, fewer electrons are reflected back at the shock (Matsukiyo
et al. 2011), so a smaller fraction of the incoming electrons are
injected in the shock-drift acceleration process. In the limit of cold
upstream plasmas studied by Riquelme \& Spitkovsky (2011), electrons
are not efficiently reflected back from the shock (black curve in
Figure~\ref{fig:time}, for $T_e=5\times10^7 K$). In this case, the
process of Weibel-mediated acceleration described above does not
operate and the resulting downstream non-thermal tail becomes steeper,
with $p\gtrsim3$ (Riquelme \& Spitkovsky 2011).

The acceleration efficiency of $5\%$ and the power-law index of $2.2$
that we find for the parameters of the bow shock of G2, combined with
our estimate for the total number of shocked electrons given in
Equation~(\ref{eq:Nsh}), allows us to write the electron energy
distribution as
\begin{equation}
\frac{dN}{d\gamma_e} \approx 2\times10^{49}\,\gamma_e^{-2.2},
\quad \gamma_e \geq 2.
\label{eq:dNdg}
\end{equation}
In the next section, we calculate the properties of the synchrotron
emission that arises from this electron distribution.

\section{Expectations for Non-thermal Synchrotron Emission}

The peak of the synchrotron emission from an electron with a Lorentz
factor $\gamma_e$ occurs at a frequency 
\begin{equation}
\nu = \frac{3}{4\pi}\gamma_e^2 \frac{eB}{m_e c}, ~~{\rm i.e.,}~~ \nughz \equiv
\frac{\nu}{10^9{\rm Hz}} \approx 2.4\times10^{-4}
\left(\frac{B}{0.06~{\rm G}}\right) \gamma_e^2, 
\end{equation}
where we have scaled the result to the expected field strength of
0.06\,G in the shocked medium. Conversely, we can invert the above
relation to infer the Lorentz factor of the electrons that contribute
predominantly to the emission at a particular frequency:
\begin{equation}
\gamma_e \approx 65 \; \left(\frac{B}{0.06~{\rm G}}\right)^{-1/2}
\nughz^{1/2}. 
\label{eq:gamma_e}
\end{equation}
The synchrotron power emitted by such electrons is 
\begin{equation}
P_{\rm synch} \approx 3.8 \times 10^{-18} \; \left(\frac{B}{0.06~{\rm
    G}}\right)^2 \gamma_e^2~{\rm erg\,s^{-1}} \approx 1.6 \times
10^{-14} \left(\frac{B}{0.06~{\rm G}}\right) \, \nughz~{\rm
  erg\,s^{-1}}.
\end{equation}
The synchrotron cooling time is then
\begin{equation}
t_{\rm cool} = \frac{\gamma_e \; m_e c^2}{P_{\rm synch}}  
\approx 6800 \; \left(\frac{B}{0.06~{\rm G}}\right)^{-2} 
\gamma_e^{-1}~{\rm yr}
\approx 105 \; \left(\frac{B}{0.06~{\rm G}}\right)^{-3/2} 
\nughz^{-1/2}~{\rm yr}.
\end{equation}
The cooling time is substantially longer than the duration of the
encounter for any value of the Lorentz factor $\gamma_e$ or,
equivalently, any frequency $\nughz$ of interest for radio or submm
observations. Thus, all the shocked electrons will contribute to the
observed synchrotron emission.

To estimate the expected spectral flux from the shocked electrons, we
use the electron Lorentz factor distribution given in 
Equation~(\ref{eq:dNdg}) and assume a distance $D=8.3$\,kpc to the 
Galactic Center. Then, using the standard formula for synchrotron 
emission from a power-law distribution of electrons 
(Rybicki \& Lightman 1979), we obtain
\begin{equation}
F_\nu \approx 19 \; \left(\frac{B}{0.06~{\rm G}}\right)^{1.6}
\nughz^{-0.6} ~{\rm Jy}, \quad p=2.2.
\end{equation}
This estimate is valid over the range of frequencies at which the
synchrotron emission is optically thin. The predicted flux is fairly
large and should be easily detected over a wide range of radio
frequencies, provided particle acceleration in the bow shock is as
efficient as the numerical simulations described in \S3 indicate.

At low frequencies, the synchrotron emission will be self-absorbed.
Using standard results (Rybicki \& Lightman 1979), we estimate the
source function of the shocked electrons to be
\begin{equation}
  S_\nu=3.7\times10^{-8} \;\left(\frac{B}{0.06~{\rm G}}\right)^{-1/2}
\nughz^{5/2} ~~{\rm erg\,cm^{-2}s^{-1}Hz^{-1}ster^{-1}}.
\end{equation}
Assuming a circular source of radius $10^{15}$\,cm at distance $D$,
the limiting synchrotron flux due to self-absorption is
\begin{equation}
F_{\nu,{\rm max}} \approx 18 \, \left(\frac{B}{0.06~{\rm G}}\right)^{-1/2}
 \nu_{\rm GHz}^{5/2}~{\rm Jy}.
\end{equation}
The quiescent emission from Sgr A$^*$ is well below this level at
frequencies above a GHz, so self-absorption should not interfere with
our ability to observe the additional emission.

\begin{figure}[t]
\centering
   \includegraphics[scale=0.43]{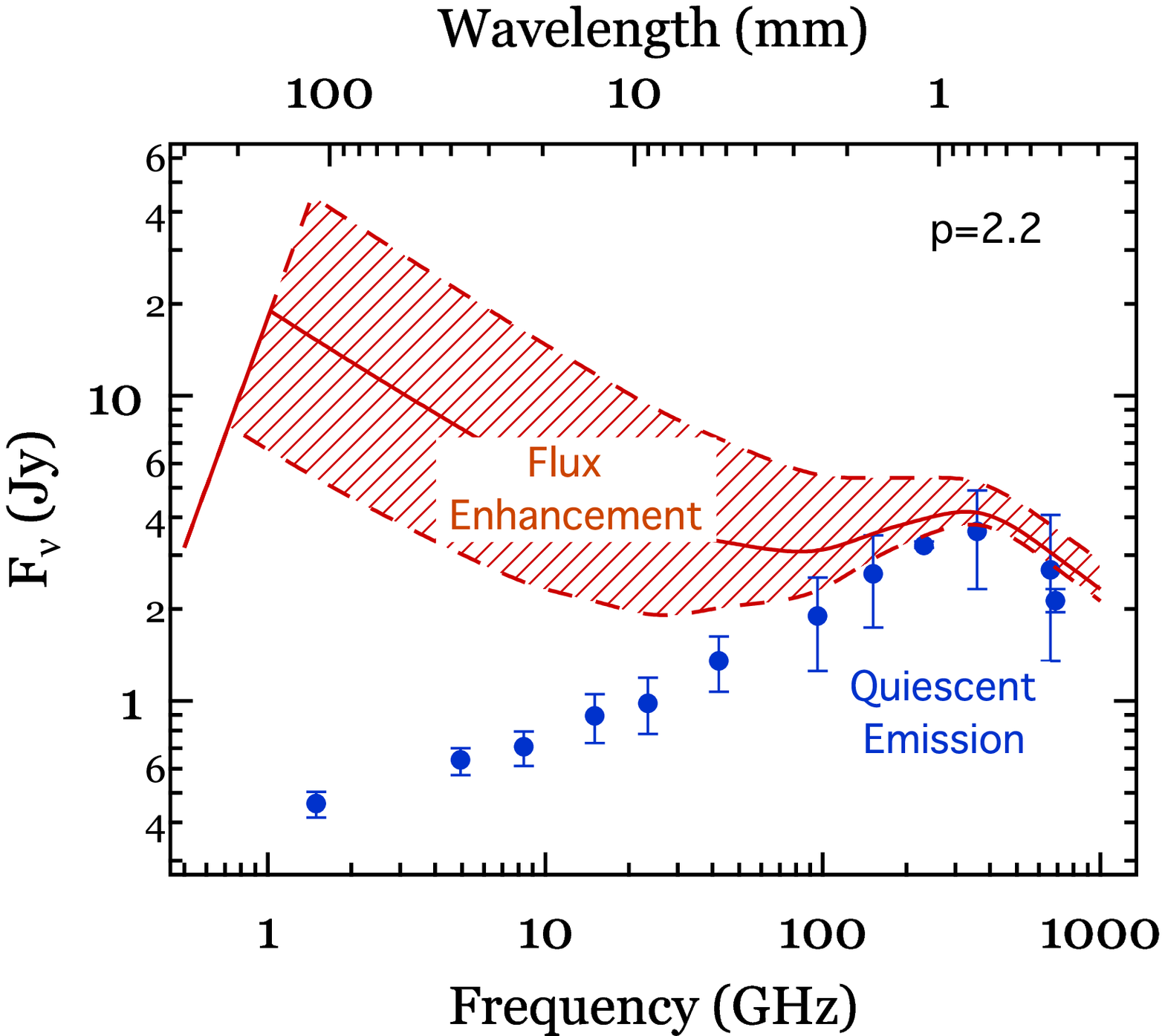}
   \includegraphics[scale=0.43]{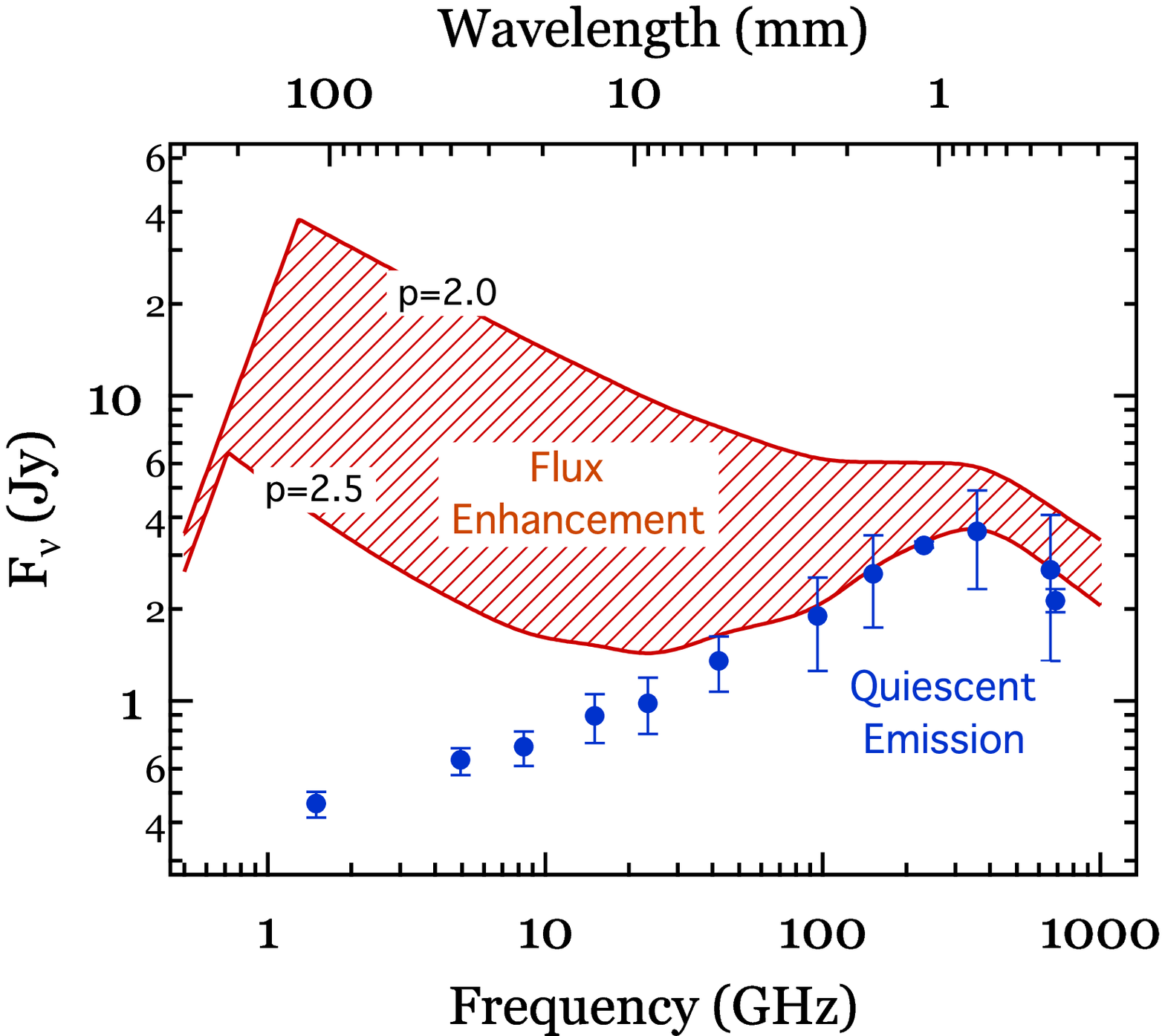}
\caption{Radio emission expected from non-thermal electrons
  accelerated in the bow shock of G2, plotted along with the quiescent
  emission observed from \sgr. The data points are compiled from
  Falcke et al.\ (1998), Zhao et al.\ (2003), and Marrone et al.\
  (2008). The solid line in the left panel shows the predicted
  spectrum for our fiducial model with power-law index $p=2.2$
  (eq.~\ref{eq:dNdg}). The hatched area bounded by dashed lines
  corresponds to a factor of three uncertainty each way in the number
  of accelerated electrons, i.e., the acceleration efficiency is
  varied from 1.7\% to 15\%. The hatched region in the right panel
  shows the predicted radio flux for power-law indices in the range
  $p=2-2.5$, keeping the fraction of accelerated electrons fixed at
  5\%. In both panels, the turnover of the flux at low frequencies is
  caused by synchrotron self-absorption. }
\mbox{}
\label{fig:sgrA_radio} 
\end{figure}

In Figure~\ref{fig:sgrA_radio}, we show the predicted radio emission
from shock-accelerated electrons in the bow shock of G2.  In order to
account for potential uncertainties in our estimates of shock
parameters (\S2) and our simulations of particle acceleration (\S3),
we explore the dependence of the expected flux enhancement on (left
panel) the number of accelerated electrons and (right panel) the
power-law index of the electron energy distribution.  We also show in
Figure~\ref{fig:sgrA_radio} the measured quiescent flux at different
frequencies, with the error bars indicating the degree of variability
among different observations. At 1.4~GHz, the quiescent radio flux of
Sgr A$^*$ is $0.5$\,Jy, whereas we estimate that the flux enhancement
could be as large as 10\,Jy. The additional synchrotron emission ought
to be easily detectable at GHz frequencies. Note also that the
spectral index is predicted to change substantially.

\section{Discussion}

The passage of the recently discovered cloud of gas G2 near
\sgr\ presents a unique opportunity to study the dynamics and
properties of hot gas in the vicinity of the black hole at the
Galactic Center.  In this Letter, we showed that a bow shock may form
during the pericentric passage of the cloud, and we investigated the
flux enhancement in the $1-100$~GHz frequency range that will arise as
a result of particle acceleration in the shock front. We ran
first-principles PIC simulations for shock parameters appropriate to
the bow shock and thereby obtained realistic estimates of the energy
distribution of accelerated electrons. Using these results, we
calculated the likely synchrotron emission from the bow shock and
found that the additional flux might exceed the quiescent emission
from \sgr\ by up to an order of magnitude at GHz frequencies. This
suggests that there is a good chance of detecting enhanced radio
emission as G2 plows through the ambient hot medium around the time of
pericentric passage. Since the cooling time of the accelerated
electrons is estimated to be long, the enhanced emission should
continue well after the encounter.

There are order unity uncertainties in the parameters we have assumed
for the bow shock, and hence the predictions made in this Letter are
not likely to be quantitatively accurate. We have allowed for some of
these uncertainties while computing the hatched regions shown in the
two panels in Figure~\ref{fig:sgrA_radio}. An additional uncertainty is
whether or not a bow shock will form in the first place. Gillessen et
al. (2012) discuss a compression shock moving into the cloud, which
will inevitably be accompanied by an external shock moving into the
ambient medium, the bow shock in our model. In most models of G2
(Burkert et al. 2012; Miralda-Escude 2012; Schartmann et al. 2012;
Murray-Clay \& Loeb 2012), the cloud retains some level of integrity
during its pericentric passage and thus is likely to develop an
external shock. However, if the cloud is completely shredded by
Kelvin-Helmholtz or other instabilities before a bow shock forms, then
our synchrotron emission estimates will no longer be valid.

Given the pericentric distance of 3100~$R_S$, which corresponds to a
projected angle of $\sim 33$~mas, the bow shock emission should be
displaced from the quiescent radio emission of \sgr\ by the same
amount. At wavelengths $\lesssim 6$~cm, this angular distance is
larger than the size of the scattering ellipse of \sgr\ (Bower et
al.\ 2006) and ought to be resolved by interferometric observations.
The scatter-broadening of the radio image of \sgr\ is believed to be
caused by a compact foreground interstellar cloud that is at least
$\sim100$\,pc from the black hole (Frail et al. 1994). Thus, the
broadening is unlikely to be affected by any gas stripped from G2
during its pericentric encounter.

In addition to the bow shock and the associated prompt radio
synchrotron emission considered in this Letter, it is expected that
the cloud G2 will also shed mass as it interacts with the ambient hot
gas. A likely early signature of the increase in the gas density at a
few thousand $R_S$ is a change in the observed Faraday rotation above
a GHz, which may provide the first estimates of the increase in the
mass accretion rate. As it moves inwards, this gas will cause the mass
accretion rate on to the central black hole to be enhanced over a
period of many years. Such an increase could cause a secular change in
the radio flux of \sgr\ on a time scale of ten years to several decades,
accompanied by changes in the ``silhouette'' of the black hole that
could be monitored by future interferometers (Moscibrodzka et
al. 2012).

\acknowledgments

R.N. thanks A. Loeb and N. Stone for useful discussions and
L.S. thanks A. Spitkovsky for insightful comments. R.N. gratefully
acknowledges support from NASA grant NNX11AE16G, and F.\"O. from NSF
grant AST-1108753. L.S. is supported by NASA through Einstein
Postdoctoral Fellowship grant number PF1-120090 awarded by the Chandra
X-ray Center, which is operated by the Smithsonian Astrophysical
Observatory for NASA under contract NAS8-03060. The simulations were
performed on the Odyssey cluster at Harvard University, on the
PICSciE-OIT High Performance Computing Center and Visualization
Laboratory at Princeton University, and on TeraGrid resources under
contract No. TG-AST120010.


\begin{thebibliography}{99}

\bibitem[{{Begelman} \& {Kirk}(1990)}]{begelman_kirk_90}
{Begelman}, M.~C. \& {Kirk}, J.~G. 1990, \apj, 353, 66

\bibitem[{{Blandford} \& {Eichler}(1987)}]{blandford_eichler_87}
{Blandford}, R. \& {Eichler}, D. 1987, \physrep, 154, 1

\bibitem[Bower et al.(2006)]{2006ApJ...648L.127B} Bower, G.~C., Goss, 
W.~M., Falcke, H., Backer, D.~C., \& Lithwick, Y.\ 2006, \apjl, 648, L127 

\bibitem[Burkert et al.(2012)]{2012ApJ...750...58B} Burkert, A., 
Schartmann, M., Alig, C., et al.\ 2012, \apj, 750, 58 

\bibitem[Falcke(1999)]{1999ASPC..186..113F} Falcke, H.\ 1999, The Central 
Parsecs of the Galaxy, 186, 113 

\bibitem[Falcke et al.(1998)]{1998ApJ...499..731F} Falcke, H., Goss, W.~M., 
Matsuo, H., et al.\ 1998, \apj, 499, 731 

\bibitem[Falcke \& Markoff(2000)]{2000A&A...362..113F} Falcke, H., \& 
Markoff, S.\ 2000, \aap, 362, 113 

\bibitem[Frail et al.(1994)]{Frail et al.} Frail, D.~A., Diamond, P.~J.,
Cordes, J.~M., van Langevelde, H.~J.\ 1994, \apj, 427, L43

\bibitem[Ghez et al.(2008)]{2008ApJ...689.1044G} Ghez, A.~M., Salim, S., 
Weinberg, N.~N., et al.\ 2008, \apj, 689, 1044 

\bibitem[Gillessen et al.(2009)]{2009ApJ...692.1075G} Gillessen, S., 
Eisenhauer, F., Trippe, S., et al.\ 2009, \apj, 692, 1075 

\bibitem[Gillessen et al.(2012)]{2012Natur.481...51G} Gillessen, S., 
Genzel, R., Fritz, T.~K., et al.\ 2012, \nat, 481, 51 

\bibitem[Macquart \& Bower(2006)]{2006ApJ...641..302M} Macquart, J.-P., 
\& Bower, G.~C.\ 2006, \apj, 641, 302 

\bibitem[Marrone et al.(2008)]{2008ApJ...682..373M} Marrone, D.~P., 
Baganoff, F.~K., Morris, M.~R., et al.\ 2008, \apj, 682, 373 

\bibitem[Matsukiyo et al.(2011)]{2011ApJ...742...47M} Matsukiyo, S., Ohira, 
Y., Yamazaki, R., \& Umeda, T.\ 2011, \apj, 742, 47 

\bibitem[{{Medvedev} \& {Loeb}(1999)}]{medvedev_loeb_99}
{Medvedev}, M.~V. \& {Loeb}, A. 1999, \apj, 526, 697

\bibitem[Miralda-Escude(2012)]{ME12}
Miralda-Escude, J.\ 2012, \apj, 756, 86

\bibitem[Moscibrodzka et al.(2012)]{MSGD12} Moscibrodzka, M.,
  Shiokawa, H., Gammie, C.~F., \& Dolence, J.~C.\ 2012, \apjl, 752, L1

\bibitem[Murray-Clay \& Loeb(2012)]{MCL12}
Murray-Clay, R.~A. \& Loeb, A.\ 2012, preprint (arXiv:1112.4822)

\bibitem[Narayan \& Yi(1995)]{NY94} Narayan, R. \& Yi, I.\ 1994,
\apj, 428, L13 

\bibitem[Narayan et al.(1995)]{1995Natur.374..623N} Narayan, R., Yi, I., 
\& Mahadevan, R.\ 1995, \nat, 374, 623 

\bibitem[Ozel et al.(2000)]{ozel00} \"Ozel, F., Psaltis, D.
\& Narayan, R.\ 2000, \apj, 541, 234 

\bibitem[Riquelme \& Spitkovsky(2011)]{2011ApJ...733...63R} Riquelme, 
M.~A., \& Spitkovsky, A.\ 2011, \apj, 733, 63 

\bibitem[Rybicki \& Lightman(1979)]{rl79} Rybicki, G.~B, \& Lightman,
  A.~P.\ 1977, Radiative Processes in Astrophysics, Wiley-Interscience

\bibitem[Schartmann et al.(2012)]{Sch12}
Schartmann, M., Burkert, A., Alig, C., et al.\ 2012, \apj, 755, 155
  
 \bibitem[{{Spitkovsky}(2005)}]{spitkovsky_05}
{Spitkovsky}, A. 2005, {in AIP Conf. Ser., Vol.~801, 345}

\bibitem[{{Weibel}(1959)}]{weibel_59}
{Weibel}, E.~S. 1959, Physical Review Letters, 2, 83

\bibitem[Xu et al.(2006)]{xu06} Xu, Y.-D., Narayan, R., Quataert, E., 
Yuan, F., \& Baganoff, F.~K.\ 2006, \apj, 640, 319 

\bibitem[Yuan et al.(2003)]{2003ApJ...598..301Y} Yuan, F., Quataert, E., 
\& Narayan, R.\ 2003, \apj, 598, 301 

\bibitem[Zhao et al.(1989)]{1989IAUS..136..535Z} Zhao, J., Ekers, R.~D., 
Goss, W.~M., Lo, K.~Y., \& Narayan, R.\ 1989, The Center of the Galaxy, 
136, 535 

\bibitem[Zhao et al.(2003)]{2003ApJ...586L..29Z} Zhao, J.-H., Young, K.~H., 
Herrnstein, R.~M., et al.\ 2003, \apjl, 586, L29 

\end{thebibliography}
\end{document}